\documentclass[12pt,aps,prd,preprint,tightenlines,superscriptaddress,
   showpacs,nofootinbib]{revtex4}
\newcommand{\PRE}[1]{{#1}} 

\usepackage{bm}
\usepackage{graphicx}

\newcommand{\gev}{\text{GeV}}

\newcommand{\etal}{{\em et al.}}

\newcommand{\eqref}[1]{Eq.~(\ref{#1})}

\newcommand{\be}{\begin{equation}}
\newcommand{\ee}{\end{equation}}

\newcommand{\bea}{\begin{eqnarray}}
\newcommand{\eea}{\end{eqnarray}}

\newcommand{\lsim}{\lower.7ex\hbox{$\;\stackrel{\textstyle<}{\sim}\;$}}
\newcommand{\gsim}{\lower.7ex\hbox{$\;\stackrel{\textstyle>}{\sim}\;$}}

\newcommand{\nbox}{{\,\lower0.9pt\vbox{\hrule \hbox{\vrule height 0.2 cm
\hskip 0.2 cm \vrule height 0.2 cm}\hrule}\,}}

\begin{document}

\preprint{UH-511-1143-09}

\title{
\PRE{\vspace*{1.5in}}
Light Dark Matter Detection Prospects at
Neutrino Experiments
\PRE{\vspace*{0.3in}}
}

\author{Jason Kumar, John G. Learned and Stefanie Smith \\%
Department of Physics and Astronomy, University of
Hawai'i, Honolulu, HI 96822, USA
}


\begin{abstract}

\PRE{\vspace*{.3in}} We consider the prospects for the detection of
relatively light dark matter through direct annihilation to
neutrinos.  We specifically focus on the detection possibilities of
water Cherenkov and liquid scintillator neutrino detection devices.
We find in particular that liquid scintillator detectors may
potentially provide excellent detection prospects for dark matter in
the $4-10\,{\rm GeV}$ mass range.  These experiments can provide
excellent corroborative checks of the DAMA/LIBRA annual modulation
signal, but may yield results for low mass dark matter in any case.
We identify important tests of the ratio of electron to muon
neutrino events (and neutrino versus anti-neutrino events), which
discriminate against background atmospheric neutrinos.
In addition, the fraction of events which arise from muon neutrinos
or anti-neutrinos ($R_{\mu}$ and $R_{\bar \mu}$) can potentially
yield information about the branching fractions of hypothetical dark matter
annihilations into different neutrino flavors.  These
results apply to neutrinos from secondary and
tertiary decays as well, but will suffer from decreased
detectability.

\end{abstract}


\maketitle

\section{Introduction}

One of the key experimental approaches to the detection
of dark matter is through indirect detection experiments.
The idea is that, in some region of elevated dark matter
density (for example, the core of the sun or the earth, or
the galactic center), dark matter particles annihilate with
each other to produce Standard Model particles which can be
detected with experiments on earth or in orbit around earth.
The focus is therefore on stable Standard Model particles,
such as $p^+$-$p^-$ pairs, $e^+$-$e^-$ pairs, photons or
neutrinos.

One possibility is that the stable particles listed above
can be produced directly through dark matter annihilation
(i.e, through the process $XX \rightarrow p^+p^-,e^+  e^-,
\gamma \gamma, \nu \bar \nu$).  The other possibility is
that dark matter particles annihilate to some other Standard
Model particles, which in turn decay by showering off the
stable particles listed above.  Either scenario has interesting
distinguishing features.  The advantage of indirect production
of stable SM particles is that it is universal;  SM particles
produced through DM annihilation will shower off at least some
set of $p,e,\gamma, \nu$ (often all of them) as they decay.
The direct annihilation of dark matter particles to stable SM
particles will, on the other hand, be suppressed by the branching
fraction to that particular final state (which may be quite small).
The advantage of direct production of stable SM states, however,
is that the SM particle is produced with an energy equal to the
dark matter mass.  This can potentially result in a sharp peak in
the energy spectrum, as opposed to the broad spectrum expected of
indirect production.

Note that we are talking about ``indirect detection" herein,
in the sense
of the neutrinos resulting from annihilations, as opposed to those
large number of ``direct detection" experiments which aim at
detecting the recoil of an elastic scattering in a laboratory
detector. This recoil typically in the KeV kinetic energy range
tells little about the nature of the WIMP, and is hard to
differentiate from background processes. Hence it seems that all
possible methods of detection will be needed to definitively
discern the existence and nature of dark matter.

One of the main theoretical candidates for dark matter has been
neutralino WIMPs, and some search strategies and analyses have been
optimized with this in mind.  However, a series of dark matter
experiments has presented hints of data which might suggest a dark
matter candidate~\cite{DMsignals}. A unifying feature of these hints
is that they are not easily explained by neutralino WIMPs.  On the
other hand, a variety of theoretical models has also arisen in
recent years which can provide reasonable dark matter candidates
which are not neutralino WIMPs, and at a wide range of masses and
couplings~\cite{nonWIMP,Feng:2008ya}. It is thus worthwhile
to revisit some of the underexplored regions of parameter space.

Our focus in this paper is on the detection of dark matter
annihilation directly to neutrinos.  This particular pathway has
not been subject to great study, partly because neutralino WIMPs
are Majorana fermions, and thus have highly suppressed
annihilations to neutrinos~\cite{Goldberg:1983nd}.
But for more general dark matter
candidates, such direct annihilations might have a significant
branching fraction.  Moreover, recent focus on leptophilic hidden
sectors \cite{ArkaniHamed:2008qn} highlight the possibility of dark
matter which annihilates primarily to leptons, in which case direct
annihilation to neutrinos may have a significant branching
fraction.  Furthermore, an advantage of direct annihilation to
neutrinos over direct annihilation to $e^+ e^-$ pairs is that the
neutrinos do not interact significantly, implying that, unlike the
case with charged SM particles, they will still provide a sharp
peak in the energy spectrum at an earth-based detector.  Another
interesting feature of neutrino signatures for dark matter
annihilation in the solar core is that the signal is largely
independent of astrophysics uncertainties.  The reason is that, for
almost all models, the sun's dark matter density is in equilibrium,
meaning that the annihilation rate is directly related to the rate
at which the sun captures dark matter.  This is in turn determined
by the dark matter mass and the dark matter-nucleon scattering
cross-section, with fewer of the uncertainties which plague other
indirect detection signatures.

The main thrust of a study of direct annihilation to
neutrinos must necessarily be focussed on the low dark matter mass
range, specifically $m_X \sim 4-10\,{\rm GeV}$.  For smaller dark
matter mass, dark matter evaporation from the sun becomes
significant~\cite{Gould:1987ju,Hooper:2008cf}, and it is difficult
to get significant bounds on the dark matter-nucleon scattering
cross-section.  For masses larger than $\sim 10\,{\rm GeV}$,
bounds on the dark matter nucleon spin-independent
scattering cross-section
from direct detection experiments already are so tight that it
appears unlikely that significant improvement can arise from
neutrino experiments.  Moreover, heavier dark matter will produce
more energetic neutrinos through direct annihilation; the muons
produced by weak interaction of these neutrinos will also be
more energetic, and much less likely to be fully-contained
within the detector (an important caveat is that higher
energy electron neutrinos may still be fully-contained, and provide
a probe of spin-dependent scattering for heavier dark matter).

Yet this narrow mass region is in itself already significant,
as dark matter in this mass range could potentially explain the
hints seen at the DAMA
experiment~\cite{Gondolo:2005hh,Bernabei:2007hw,Petriello:2008jj,Savage:2008er,Chang:2008xa}
(though see also~\cite{Fairbairn:2008gz}). This range of light dark
matter has been the subject of much recent theoretical and
phenomenological
interest~\cite{Bottino:2003iu,Feng:2008ya,Feng:2008dz,DAMADMmodel}
and there has already been significant interest in the possibility
of investigating this range with the Super-Kamiokande
experiment~\cite{Desai:2004pq,Hooper:2008cf,Feng:2008qn,Andreas:2009hj}
(see also~\cite{Kusenko:2009iz}).
Hence it is worth seeing what can be done to improve sensitivity to
dark matter in this mass range.

In section 2 we will review the types of neutrino detection
experiments which are relevant for this discussion, in particular
water Cherenkov and liquid scintillator detectors.  In section 3 we
will review the possibility of using liquid scintillator detectors
to provide directionality information on incoming neutrinos, which
allows for much greater sensitivity to dark matter annihilations in
the sun.
In section 4 we will discuss the relationship between observed
electron neutrinos and muon neutrinos, and the implications for
dark matter annihilation.
In section 5 we will exhibit the types of bounds which
current and future neutrino experiments can obtain for light dark
matter which annihilates directly to neutrinos.  We conclude with a
discussion of our results in section 6.

\section{Detector Overview}

Clearly to make progress in indirect dark matter sensing via
neutrinos, we need ever larger instruments, and instruments with
better resolution for the presently considered mono-energetic
neutrinos from dark matter.  The present largest water Cherenkov
detector, Super-Kamiokande (SK)~\cite{Ashie:2005ik},
has a fiducial volume
of 22,500 m$^3$. There are several proposals for larger
instruments, such as Hyper-Kamiokande, UNO, Memphys, and possible
instruments at the proposed DUSEL in the Homestake mine.  All of
these are in the class of 20-50 times larger in mass than SK.

For Cherenkov instruments, neutrino signals for the energy range
from a few tens of MeV upwards are dominated by neutrinos
generated by cosmic rays striking the atmosphere. At energies of
the order of one to a few GeV the ratio of muon neutrinos to
electron neutrinos is roughly equal, as it would have been 2:1
except half the muon neutrinos have equilibrated with tau neutrinos
through oscillations (and at terrestrial distances the electron
neutrinos of these energies have not oscillated much at all).
The quasi-elastic neutrino interaction dominates at these low
energies, and well developed algorithms distinguish between
electron and muon events with 99\% efficiency.  The energies of
individual events are measured to a few percent and angles to
several degrees.  There is an inherent coupling between vertex
resolution and energy and angle in these ring measuring detectors,
which prevents reaching the better resolutions one might expect
due to hundreds or even thousands of phototube hits.

Liquid scintillation detectors, of which KamLAND is the largest in
existence with a 600 ton fiducial mass, have far greater light
production per unit energy deposition (for example,
250 photoelectrons (PE)/MeV at KamLAND versus 10 PE/MeV at SK).
Borexino, at 100 tons, also contributes.  These will soon be joined
SNO+, a liquid scintillator detector in Canada with a similar target mass.
Other detectors are under discussion and proposal: the portable
deep ocean 10 kiloton Hanohano instrument, the 50 kiloton LENA
instrument (in a mine cavity in Europe), and potential detectors in
Homestake.  These instruments have been designed largely for
detecting geo-neutrinos and reactor neutrinos, as well as for
detecting proton decay and the diffuse supernova neutrino background.

For liquid scintillation detectors, the advantage of greater light
production (compared to the highly directional
Cherenkov radiation) is offset by the uniformity of radiation of the
scintillation light.  This has been thought, until recently, to
obviate the use of liquid scintillation detectors for directional
information or for neutrino flavor identification. As
suggested in \cite{Learned:2009rv} it may be possible to employ the times
of the first hits at each photomultiplier tube (PMT) to define a
``Fermat surface" which can be back-projected in a type of
tomography to make good reconstruction of simple particle
topologies (as from atmospheric neutrino events and nucleon
decay), including strong flavor identification. The simple Monte
Carlo program results referred to, at this time indicate
resolutions in angle and energy perhaps ten times smaller than for
SK (and presumably also for future water Cherenkov detectors).
We do not wish to make potentially controversial claims herein, but
only to note the importance of such sensitivity to dark matter
detection, as we discuss below.  We will thus take the water
Cherenkov resolution as 3\% in muon energy and 3 degrees in angle,
scaling with the square root of energy in GeV on an event by event
basis~\cite{Ashie:2005ik}.  The
optimistic case for scintillators we take as 10 times
better in each parameter.  The real world is probably somewhere in
between.

\section{Directionality and Determining Neutrino Energy}

Water Cherenkov detectors determine the direction of a muon or an
electron event record via fitting of ``Cherenkov rings'' to the PMT
hits (employing both time and amplitude). The liquid scintillation
detectors can utilize the Fermat surface to do the same, except
that the light from near the beginning of the track and near the
end of the track give point source signatures that well define the
track vertex and end point.

For neutrinos coming from the sun (or earth
center or galactic center), we know (assume) the incoming neutrino
direction, calculable at the detector at any moment.  Measuring the
relativistic muon momentum relative to this direction, yields the neutrino
energy:
\bea
E_{\nu} &\approx& {m_N E_{\mu} \over
m_N -E_{\mu} (1-\cos \theta)}.
\eea
Given estimates of measurement uncertainties, plus the angular
distribution expected, the neutrino energy (and hence dark matter
mass) resolution
of water Cherenkov detectors in the 4-10 GeV range
will be $\sim 3-20\% $, depending
on the angle of the muon with respect to the sun.  An
average estimate of the energy resolution is 10\% for
water Cherenkov detectors, and up to 1\% for
optimistic future liquid scintillation detectors.

\section{Importance of the Observed Flavor Ratio}

Note that most discussions have focussed upon muon neutrino
detection.  However, electron neutrinos are very useful here since
they are fully contained more readily in the detectors (hence
larger effective target volumes, particularly at energies extending
to around 100 GeV, compared to a few GeV for containing muon
events). No matter what the source annihilation neutrino fraction,
neutrino mixing will deliver a mixed beam at our detectors. The
technology for discriminating electron events from muon events in
the energy range of around 1 GeV is very well developed and gives
separations to order of a percent crossover, which for our case
here would seem to be more than adequate.

Detection of dark matter muon neutrinos requires consistent
detection of electron neutrinos and vice versa due to inescapable
neutrino oscillations.
There are two important regimes to note: neutrino oscillations
within the sun and neutrino oscillations in the vacuum.
Neutrinos produced by dark matter annihilation in the sun are
produced at the core, which is effectively a point source.  As
these neutrinos pass through the sun, electron neutrinos will
scatter off background electrons through W-boson mediated
interactions, while
other neutrinos will not.  This interaction will significantly
modify neutrino oscillation within the sun~\cite{MSW}.
These effects
have been calculated in detail
in~\cite{Lehnert:2007fv} (note that neutrinos and anti-neutrinos
will have different matter-induced oscillations).
Neutrinos leaving the sun will subsequently
exhibit vacuum oscillation as they travel to the earth.
In the case of a pure $\nu_e$ source emerging from the sun, the
ratios after the flight time for mixing will be
$e$/$\mu$/$\tau$ = 5/2/2, and for a pure $\nu_{\mu}$ source it will
be 4/7/7.  For the equal production of $\nu_e$ and $\nu_{\mu}$ with
little $\nu_{\tau}$ (for example, due to low mass of the dark
matter), we should see the ratios at earth as 14/11/11
~\cite{Pakvasa:2004hu}.

In all these cases the $\tau$ appearance will be
slight and difficult to resolve, so we will have to rely upon
$\mu$ to $e$ ratios to untangle the dark matter physics.
Presently conceived detectors cannot distinguish between
$\nu_e$ and $\bar{\nu_e}$ at these energies (perhaps later a liquid
argon instrument with magnetic field can do this).  But these
detectors can distinguish, statistically, between $\nu_{\mu}$ and
$\bar{\nu_{\mu}}$ due to stopped muon decay ($\mu^-$ generally get
absorbed on nuclei and $\mu^+$ decay, detectably).
In the case of a magnetized iron detector such as the proposed INO,
the sign of the $\mu^{\pm}$ charge can be measured directly by the
curvature of the muon track.
Though the $\nu_{\mu}$ and the $\nu_{\bar \mu}$
have different cross-sections in the slightly isospin
asymmetric target material (oil or water) and the y-distributions
are different, the event ratios should be predictable to a few
percent. Moreover these rates will be somewhat different for dark
matter neutrinos and the background atmospheric neutrinos, providing
another signature for dark matter neutrinos.
Thus, the observables we use for this flavor analysis are
$R_{\mu} \equiv {N_{\mu} \over (1/2)(N_{\mu}+N_{\bar \mu} +N_{e,\bar e})}$ and
$R_{\bar \mu} \equiv {N_{\bar \mu} \over
(1/2)(N_{\mu}+N_{\bar \mu} +N_{e,\bar e})}$,
which are the ratios of the number of
muon (or anti-muon, respectively) events to the total number of
$\mu^\mp$, $e^\mp$ neutrino events.

\begin{figure}
\resizebox{5.0in}{!}{
\includegraphics{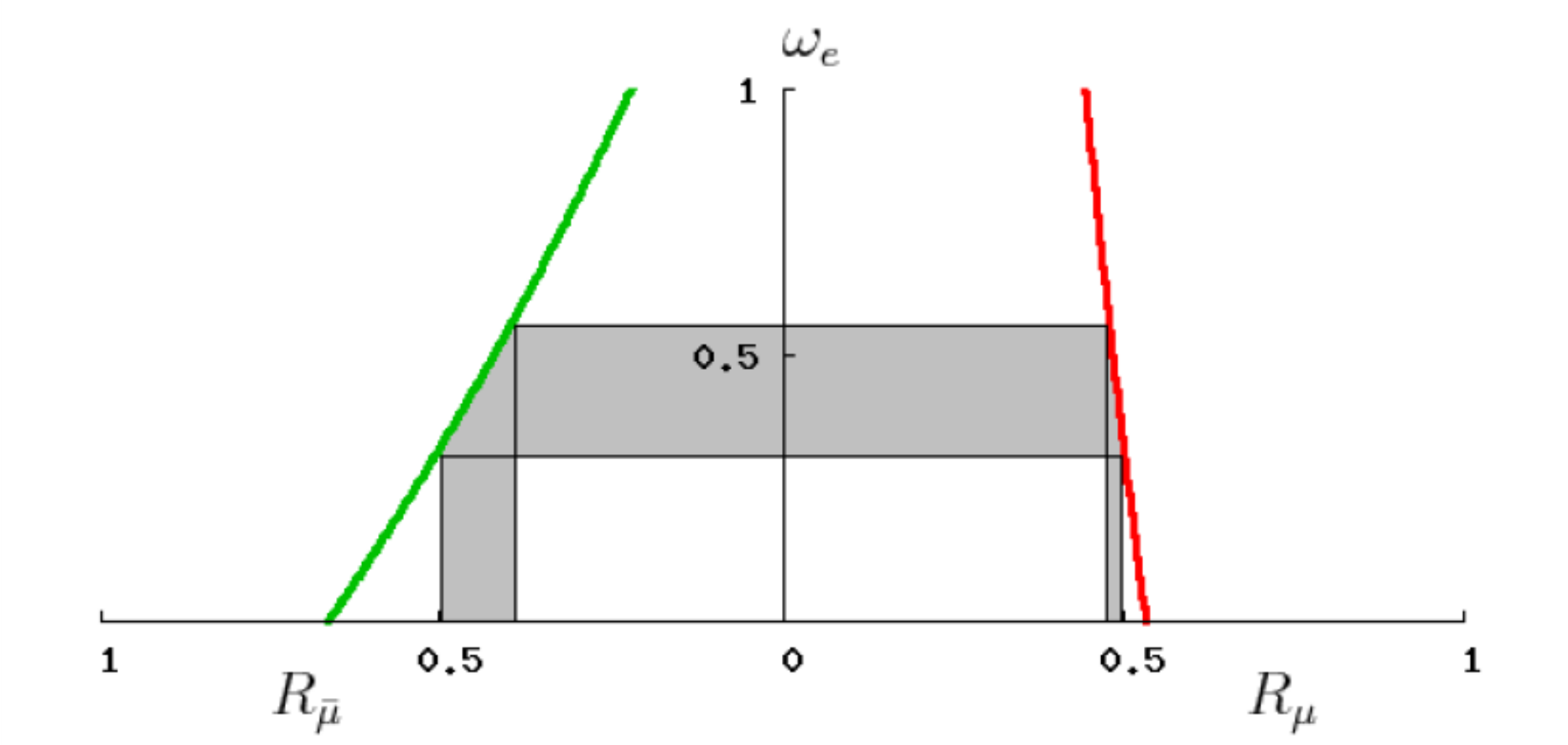}
}
\caption{The electron neutrino fraction produced from dark matter annihilation,
as a function of $R_{\mu}$ and $R_{\bar \mu}$.  The shaded region corresponds
to values of $w_e$ and $R_{\mu}$ consistent with $R_{\bar \mu} = 0.44 \pm 0.04$.
Note that only the regions ${4\over 9} \leq R_{\mu} \leq {8\over 15}$ and
${2\over 9} \leq R_{\bar \mu} \leq {2\over 3}$ are consistent with dark matter
annihilation in the sun with tribimaximal mixing and $\theta_{13}=0$.
\label{fig:hanofig2}
}
\end{figure}

We assume that lepton flavor is conserved
in dark matter annihilation.
If one further assumes tribimaximal mixing (which is quite consistent
with experimental data), and $\theta_{13} =0$, then
the oscillation
matrix (through both the sun and vacuum) is entirely determined by the
$w_{e}$, the fraction of (anti-)neutrinos produced at the sun's
core by dark matter
annihilation which are of electron type~\cite{Lehnert:2007fv}.
In particular, one then finds that ${1\over 3}$ of all neutrinos
arriving at earth (after vacuum and matter-induced oscillations) are
of $\mu$-type, while the fraction of anti-neutrinos of $\mu$-type
arriving at earth is given by ${1\over 12} (5-3w_e)$.  Since
we have two observables and one parameter, we find two
independent determinations of $w_e$ (see figure \ref{fig:hanofig2}):
\bea
w_e &=& {5\over 3} \left( {2- 3R_{\bar \mu} \over 2+ R_{\bar \mu}} \right),
\nonumber\\
&=& {8\over 3R_{\mu}} -5.
\eea

As we see from the figure, the constraint $0 \leq w_e \leq 1$ imposes
two independent consistency conditions on the muon fraction observables.
An inconsistency in the measurement of either would falsify the ansatz
of flavor-conserving dark matter annihilation, with tribimaximal mixing
and $\theta_{13}=0$.  For the ansatz $\theta_{13}\neq 0$, a similar analysis
may be performed along the lines of~\cite{Lehnert:2007fv}; this analysis
is more complicated and beyond the scope of this work.

Note that range of $R_{\mu}$ consistent with the above ansatz is
quite narrow; to make a measurement of $R_{\mu}$ of greater precision
would require an unrealistic number of events.  But the range of
$R_{\bar \mu}$ consistent with this ansatz is larger, and with
perhaps 500 total $\nu$, $\bar \nu$ events one might conceivably obtain
a measurement of $R_{\bar \mu}$ with 10\% accuracy within this range.

\section{Dark Matter Bounds from Neutrinos}

Dark matter accumulates in the core of the sun due to elastic
capture scattering from solar nuclei.  If a dark matter particle
loses enough energy to nuclear recoil, its velocity will fall below
the escape velocity.  Dark matter is then gravitationally captured
by the sun, and eventually settles at the core. The capture rate,
$\Gamma_C$, is
thus largely determined by the dark matter-nucleon scattering
cross-section and the dark matter number density.

In particular, we find
\bea
\Gamma_C \sim \left( {\sigma_{X-N} \over m_X } \right)
\left( 2\times 10^{29} {\gev \over {\rm pb}\,{\rm s}} \right).
\eea
The coefficient\cite{Feng:2008qn,Hooper:2008cf} is
accurate up to ${\cal O}(1)$ factors related to composition
and motion of the sun, the halo density, etc.,
but it will be sufficient for our purposes.

It was shown in \cite{Gould:1987ir,equilib,Hooper:2008cf}
that for the range of dark-matter masses of
interest, the sun is in equilibrium (a similar result has been shown
for the case where dark matter primarily interacts with electrons
at tree-level \cite{Kopp:2009et}).  This implies that
$\Gamma_C = 2\Gamma_A$, where $\Gamma_A$ is the total dark matter
annihilation rate.    In the case of the direct annihilation (which we
assume here), $\Gamma_{\nu_{\mu},\bar \nu_{\mu}}
= {2\over 3} B_{\nu} \Gamma_A $, where
$B_{\nu}$ is the branching fraction to all neutrino species
(the factor of ${1\over 3}$ is a rough assumption from neutrino oscillation; the
actual value depends on the initial flavor ratios as discussed above).
It is this neutrino production rate which we can bound
with limits from neutrino experiments.

We are interested here in the event rate for fully-contained muons
(i.e., muons which are created by a neutrino weak interaction
within the detector and which stop within the detector).  These are
of interest, because it is only for fully-contained events
that we can measure the total energy of the neutrino, which
exhibits a peak in the energy spectrum.

We define $N$ as the number of events needed for a discovery of
dark matter after a run-time $T$.  This corresponds to a solar
neutrino production rate due to dark matter of
\bea
\Gamma_{\nu_{\mu}, \bar \nu_{\mu}} = {N\over T}
{ 4\pi(1.5\times 10^{11} {\rm m})^2 \over \sigma_{FC}^{eff.}},
\eea
where $\sigma_{FC}^{eff.}$ is the effective cross-section for
the detector to produce fully-contained muon events.
This effective cross-section is given by
\bea
\sigma_{FC}^{eff.} &=& \sigma_{\nu-N} \times
{\rho \over m_N} \times {\rm eff.\,volume},
\eea
where $\rho$ is the density of the detector and the effective
volume is the approximate volume of the detector which can yield
fully-contained muons for the given neutrino energy (this effective
volume is dependent on the detector geometry).  Here,
$\sigma_{\nu-N}$ is the neutrino(anti-neutrino)-nucleon scattering
cross-section.

For the energy range of interest, $M_W \gg E_{\nu} > m_N$, where
$m_N$ is the mass of a nucleon.  In this case, the
(anti-)neutrino-nucleon scattering cross-section can be
approximated as~\cite{Gandhi:1995tf,Jungman:1995df,Edsjo:1997hp}

\bea
\sigma_{\nu n} &=& 8.81 \times 10^{-3}\,{\rm pb}\,
(E_{\nu}/{\rm GeV}),
\nonumber\\
\sigma_{\nu p} &=& 4.51 \times 10^{-3}\,{\rm pb}\,
(E_{\nu}/{\rm GeV}),
\nonumber\\
\sigma_{\bar \nu n} &=& 2.50 \times 10^{-3}\,{\rm pb}\,
(E_{\nu}/{\rm GeV}),
\nonumber\\
\sigma_{\bar \nu p} &=& 3.99 \times 10^{-3}\,{\rm pb}\,
(E_{\nu}/{\rm GeV}).
\eea
One should also include the small contribution from
resonant, coherent and diffractive processes, but this
approximation will be sufficient for our purposes.

For specificity, we can use Super-Kamiokande as an example
(see also \cite{Desai:2004pq}).
We find there that
\bea
\sigma_{FC}^{eff} \sim
(1.4 \times 10^{-8}) \left( E_{\nu} \over GeV \right)
({\rm meter})^2
\times \left( {\rho \over {\rm g}/{\rm cm}^3} \right)
\times {\rm vol.\,factor},
\eea
where ``vol. factor" is the factor by which the effective
volume of the detector in question for fully-contained events at the
given energy exceeds the fiducial volume of SK.  This gives us
a detection limit

\bea
\label{maineq}
\sigma_{XN} \simeq \left( {1.2\times 10^{-3} {\rm pb}
\over {\rm vol.\,factor}} \right)
\left( {\rho \over {\rm g}/{\rm cm}^3} \right)^{-1}
\left( {3 \over \sum_F B_F \langle Nz \rangle_F  }\right)
\left( {N_{events} \over N_{live-days}} \right),
\eea
where $z = {E_{\nu} \over m_X}$.
Note that one achieves the same bound for fully-contained
electrons arising from electron neutrinos interacting with
the detector.

\subsection{Limits for a liquid scintillator detector}

We are now ready to put the pieces together to obtain a basic
analysis of the detection prospects for dark matter at a
liquid scintillator neutrino detector via the annihilation process
$XX \rightarrow \nu \bar \nu$ in the sun's core.

The key point here is that a liquid scintillator can be expected to
provide a high resolution measurement of the full neutrino energy
for interactions which produce a fully-contained muon.  Of course,
there are efficiencies which depend on the details of the
scattering process (such as whether it is best characterized as
quasi-elastic, resonant or deep-inelastic) and the details of the
detector.

Dark matter annihilation in the sun produces neutrinos which
arrive at earth from a known direction.
However, scattering within
the detector will produce leptons in a cone around the direction
to the sun, with rms half-angle $\theta \sim 20^\circ
\sqrt{10\,{\rm GeV} /E_{\nu}}$~\cite{Jungman:1995df}.
The background to this signal
would be fully-contained muons arising from interactions of
atmospheric neutrinos with the detector.
The Super-K collaboration reports~\cite{Ashie:2005ik} that,
for $E_{\nu_{\mu}} \gsim 4\,{\rm GeV}$,
they expect less than 20 fully-contained muon
events (in the cone of the sun)
per GeV neutrino energy bin per 1000 live days running time
due to atmospheric neutrinos
(the rate falls to 2 events per GeV per 1000 live days at
$E_{\nu_{\mu}} \sim 10\,{\rm GeV}$).
Using the fiducial volume of Super-K (V = 22,500 $m^3$) and a
$N_{live-days} = 3000$  runtime as a guide, we would thus expect
$\lsim 1$ fully-contained muon background
event per energy bin (assuming a
1\% energy resolution).
We are thus in the limit of small statistics; a detection of
10 fully-contained muon events in the cone of the sun, with total
energy in the same bin should be sufficient to
detect dark matter annihilation in the sun with $m_X$ given by the
measured neutrino energy.

\begin{figure}
\resizebox{5.0in}{!}{
\includegraphics{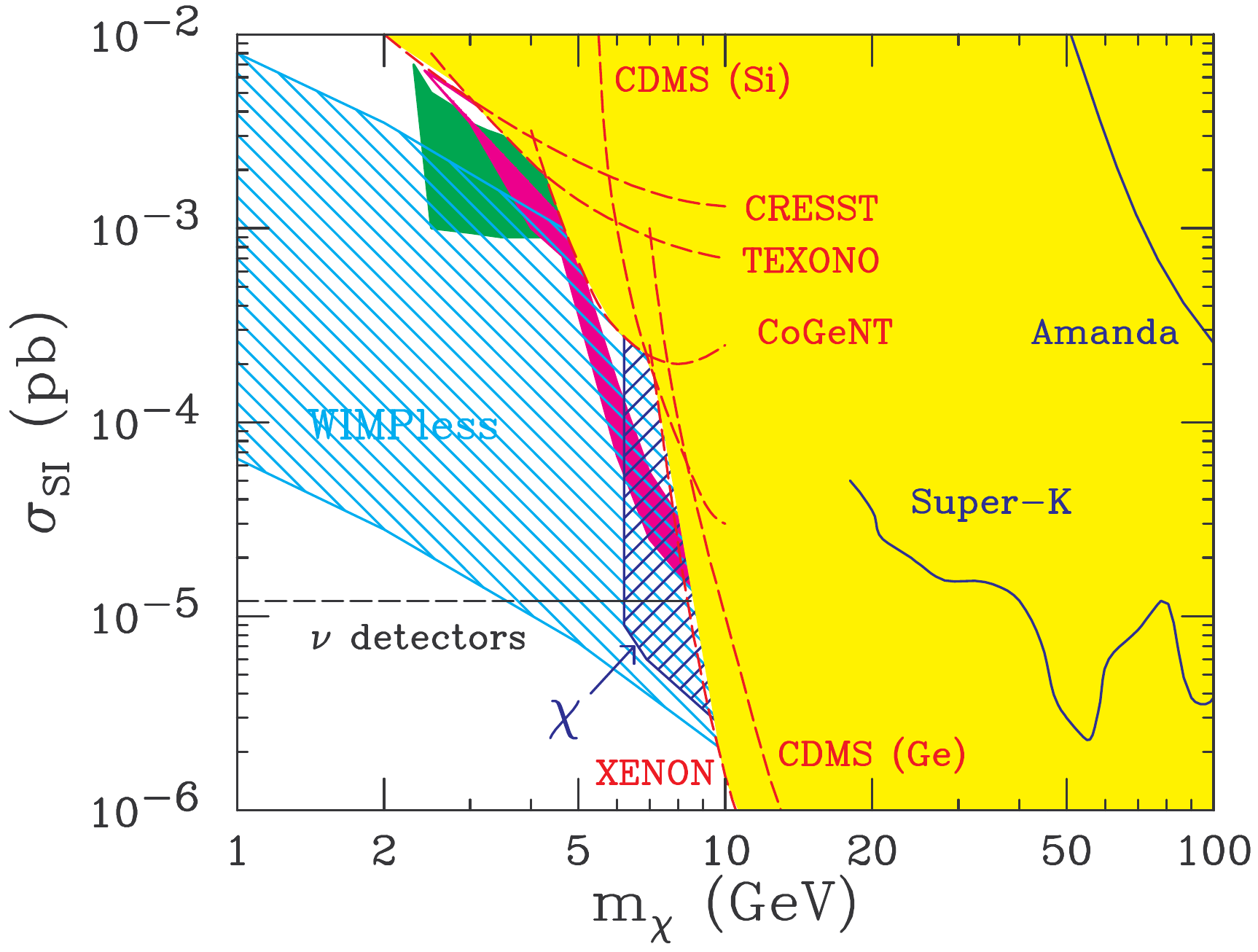}
}
\caption{Bounds on spin-independent
$X$-nucleon scattering as a function of dark matter mass $m_X$.
The magenta shaded region is
DAMA-favored given channeling and no streams~\cite{Petriello:2008jj}, and
the medium green shaded region is DAMA-favored at 3$\sigma$ given
streams but no channeling~\cite{Gondolo:2005hh}.  The light yellow
shaded region is excluded by the direct detection experiments
indicated ~\cite{directdetectlimits}.  The dark blue
cross-hatched region is the prediction for
the neutralino models considered in Ref.~\cite{Bottino:2003iu} and the
light blue slashed region is the parameter space of a class of
WIMPless models considered in~\cite{Feng:2008qn,Feng:2008ya}.
The indicated blue solid lines are the published limits from
SK~\cite{Desai:2004pq} and AMANDA~\cite{Braun:2009fr} (assuming annihilation
to $b \bar b$).
The black solid line is the detection threshold for liquid scintillator
neutrino detectors of $\rho = 1\,{\rm g}/{\rm cm}^3$ with
$22,500 {\rm m}^3$ fiducial volume running for
3000 live days, assuming annihilation only to neutrinos and
detection only of $\nu_{\mu}$. The black dashed line indicates the
sensitivity of liquid scintillator neutrino detectors if
WIMP evaporation effects are ignored.
\label{fig:hanofig}
}
\end{figure}

Furthermore, for direct annihilation to neutrinos, we have
$E_{\nu} = m_X$, so we may take $\langle Nz \rangle =1$.  We thus
find:
\bea
\sigma_{XN} \simeq  1.2\times 10^{-3} {\rm pb} \times
\left( {\rho \over {\rm g}/{\rm cm}^3} \right)^{-1}
\left( {3 \over B_{\nu}   }\right)
\left( {N_{events} \over N_{live-days}} \right)
\left( {22,500 m^3} \over {V} \right),
\eea
and our bound is largely independent of $m_X$.
Choosing $\rho = 1\,{\rm g}/{\rm cm}^3$,
$N=10$, $N_{live-days}=3000$, $V=22,500 m^3$
and $B_{\nu}=1$ yields the
discovery potential plotted in figure \ref{fig:hanofig}.
Note that this bound is based only on detection prospects from
$\nu_{\mu}$,$\bar \nu_{\mu}$.  A similar bound would result if one
only studied the $\nu_{e}$,$\bar \nu_{e}$ signal.  A combined
analysis of both signals will improve detection prospects by
approximately a factor of 2.  But a more detailed analysis would
be required to account for varying efficiencies for each of the two
signals, and is beyond the scope of this paper.

It is worth noting that neutrino detectors will produce similar
bounds on the spin-dependent dark matter-proton scattering
cross-section, due to capture from hydrogen.  This is
of interest because direct detection bounds on spin dependent
dark matter-nucleon scattering cross-sections are much weaker
than in the spin-independent case.  It might thus be worthwhile
to extend this analysis to dark matter masses greater than
$10\,{\rm GeV}$, where the sensitivity found in this analysis may beat
current spin-dependent bounds (remember, the bound from direct
annihilation is largely independent of the dark matter mass).

\begin{figure}
\resizebox{5.0in}{!}{
\includegraphics{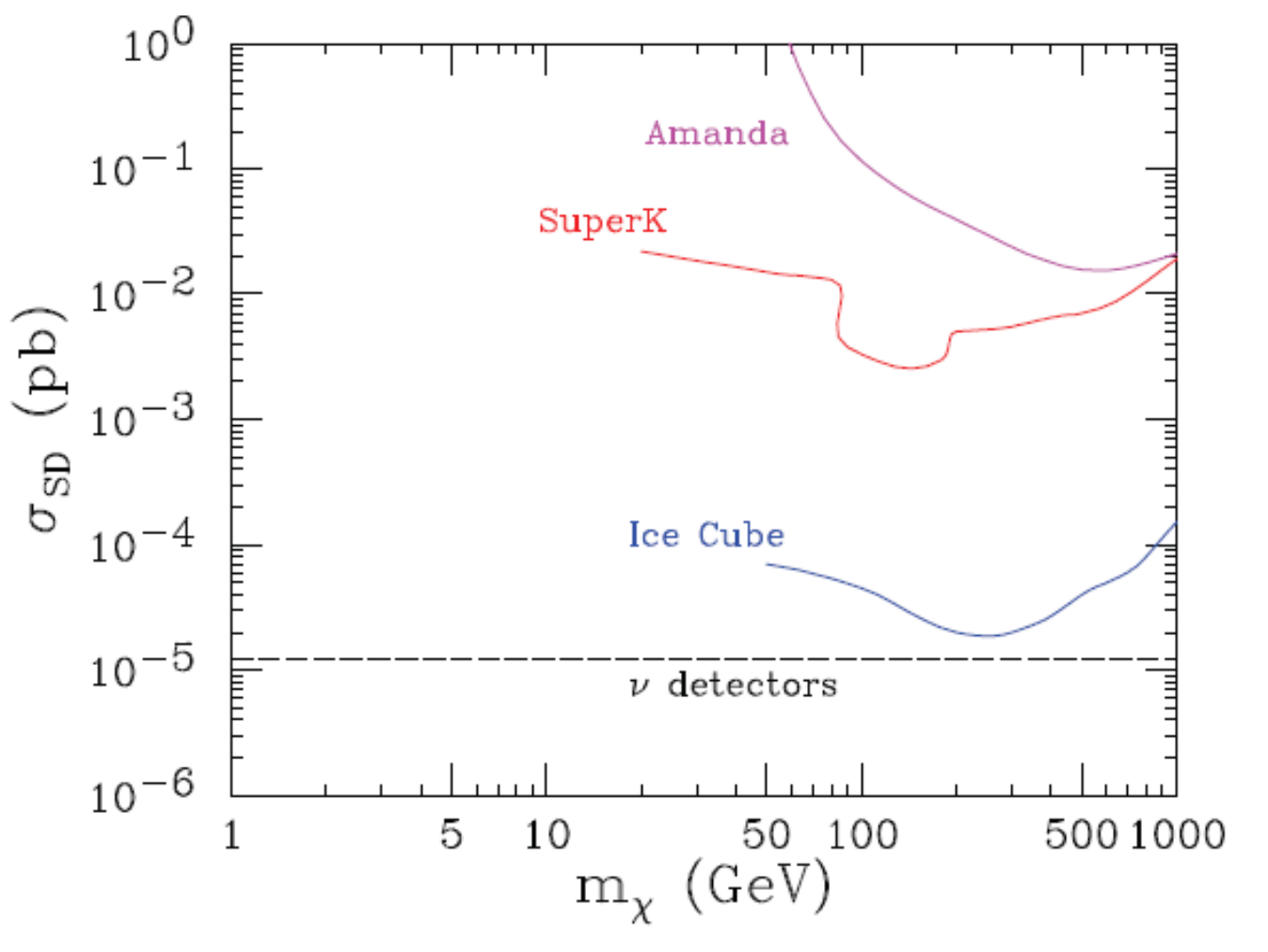}
}
\caption{Bounds on spin-dependent
$X$-proton scattering as a function of dark matter mass $m_X$.
The indicated red line is the published limit from
SK~\cite{Desai:2004pq} and the magenta line is the published
limit from AMANDA~\cite{Braun:2009fr} (assuming annihilation
to $b \bar b$, and the analysis of 2001-2003 data).  The blue line
is the projected limit from IceCube-80+DeepCore~\cite{Braun:2009fr}
(assuming 1800 live days).
The dashed black line is the detection threshold for liquid scintillator
neutrino detectors of $\rho = 1\,{\rm g}/{\rm cm}^3$ with
$22,500 {\rm m}^3$ fiducial volume running for
3000 live days, assuming annihilation only to neutrinos and
detection only of $\nu_{e}$.
\label{fig:hanofigSD}
}
\end{figure}

However, we are still limited by the fact that we need
fully-contained leptons.  For $m_X > 10\,{\rm GeV}$, even
a detector significantly larger than SK will still see few
fully-contained muons.  However, the electrons produced from
$\nu_{e}$ interactions will be fully-contained.  An
analysis of these fully-contained electron events may potentially allow
one to extend the analysis described here
to $m_X \gg 10\,{\rm GeV}$ in the case of spin-dependent
scattering.
Assuming similar energy and angular resolution for electrons and
muons, we have estimated the detection prospects for the type of
detector described above in Fig. \ref{fig:hanofigSD}.
However,
a more accurate bound requires a detailed treatment of the scintillator's
response to the electron showers.
A more detailed analysis of neutrino detector
bounds from electron events seems to be a
very promising avenue for further investigation.

\subsection{Limits for a water Cherenkov detector}

For a water Cherenkov detector, the main difference in detection
prospects will be in the
energy (mass) resolution and in the
number of events $N$ needed for detection.
The measurement of the energy and angle of the
muon will yield as sharp a peak in reconstructed neutrino energy
distribution.  But the neutrino energy resolution we expect is only
$\sim 10 \%$.  So for neutrino energies $\sim 10\,{\rm GeV}$
we still expect to be in the limit of small statistics, and the
previous bounds hold.  But for $E_{\nu} \sim 4\,{\rm GeV}$, one would
expect close to 40 atmospheric neutrino events per energy bin
(assuming $10\%$ energy resolution)~\cite{Ashie:2005ik}.  In this
regime, demanding $\rm{signal} / \sqrt{\rm{background}} =5$ for the discovery
of an excess over known background should require about 30
events\footnote{The uncertainty in the measurement of the excess is
determined by $\sqrt{\rm{signal + background}}$.}.  The
sensitivity of a water Cherenkov detector would thus be about a factor
of 3 worse (for a run-time of 3000 live-days and a volume of
$22,500\,{\rm m^3}$).

The important point for us is to note the scaling of our limit with
detector volume and with run-time.  For the cases where small statistics
are relevant, we found that our detection limit scaled
inversely with both volume and run-time.
But when background becomes
significant, our sensitivity will scale as
$\rm{signal} /\sqrt{\rm{background}} \propto
\sqrt{\rm{volume} \times \rm{time}}$.
Our detection bound will thus scale as
$(\rm{volume} \times \rm{time})^{-{1\over 2}}$.
One should
note that the limits obtained here for both liquid
scintillator and water Cherenkov detectors are dependent on
${\cal O}(1)$ factors related to both astrophysics and the
particle physics of the sun and the detector response.

\subsection{Other Dark Matter Annihilation Modes}

Having done the case of leptophilic neutrino annihilations, we are
in a position to generalize these results for more general decays
into various combinations of decay products, as discussed earlier.
One may think of the leptophilic case as the Green's function,
which must be swept over the decay spectra for other types on
annihilation products. Detectability of dark matter suffers as the
signal to noise gets worse, naturally.  But the issues of flavor
ratios remain, for each energy.  So, while the signal to noise will
take a beating, the information on the dark matter decays will
increase, if resolvable, potentially revealing whether the
neutrinos are primary, secondary or even tertiary decay products.
Clearly some work is needed in this area, which we have not yet
completed.

\section{Conclusions}

We have investigated the prospects for liquid scintillator-type and
water Cherenkov-type neutrino detectors to discover dark matter
through the $XX \rightarrow \nu \bar \nu$ annihilation process in
the core of the sun.  We have found that the high energy-resolution
available at a liquid scintillator detector, combined with its
ability to resolve the directionality of muons, would give a liquid
scintillator detector excellent detection prospects in the
$4-10\,\gev$ range.  Although a water Cherenkov detector will also
have very good prospects in this range, its (presumably) lesser energy
and track angle resolution gives an advantage to liquid
scintillator-type detectors (assuming early calculations for these
detectors are indeed realizable).  In any event, we have pointed out
that the coupled detection of both muon and electron neutrinos from
annihilations will yield important and unique information of the
dark matter branching fractions.  We have identified a parametrization
of the ratios of muon and anti-muon events to the total which can
reveal the source $\nu_{e, \bar e}$ fraction in dark matter annihilations,
yielding perhaps the first detailed measurement of the
structure of dark matter interactions.

This light dark-matter range is of particular interest, since
it is a mass range which can potentially explain the DAMA
result.  Indeed, there has been much recent interest in
leptophilic dark matter candidates, for which the annihilation
channel to neutrinos can be significant.  For this possibility,
low threshold neutrino experiments should provide the best
bounds on the dark matter-nucleon scattering cross-section.
In particular, we find that with significant branching fraction to 
$\nu \bar \nu$, low-threshold neutrino experiments should probe much 
of the low-mass parameter space consistent with the dark matter 
interpretation of DAMA.

For heavier dark matter, it is possible that an analysis of
fully-contained electron events can yield significant detection
prospects.  Interestingly, it has been recently argued that
neutrino probes of dark matter can also reveal information about
local dark matter density fluctuations in the regions through
which the sun has traversed in the past~\cite{Koushiappas:2009ee}.
Moreover, for models in which dark matter self-scattering is
enhanced, the neutrino flux from dark matter annihilation may be
significantly enhanced~\cite{Zentner:2009is}.
It seems that dark matter detection at neutrino detectors may
have a very bright future.

\section*{Acknowledgments}

We are grateful to M. Batygov, S. Bornhauser, B. Dutta, S. Dye,
J. Feng, S. Pakvasa, K. Richardson-McDaniel
and X. Tata for useful discussions.
This work is supported in part by Department of Energy grant
DE-FG02-04ER41291.

\end{document}